# Cluster magnetic octupole induced out-of-plane spin polarization in antiperovskite antiferromagnet

Yunfeng You[1], Hua Bai[1], Xiaoyu Feng[2], Xiaolong Fan[2], Lei Han[1], Xiaofeng Zhou[1], Yongjian Zhou[1], Ruiqi Zhang[1], Tongjin Chen[1], Feng Pan[1] & Cheng Song[1,*]

[1]Key Laboratory of Advanced Materials, School of Materials Science and Engineering, Tsinghua University, Beijing 100084, China

[2]The Key Lab for Magnetism and Magnetic Materials of Ministry of Education, Lanzhou University, Lanzhou 730000, China

*songcheng@mail.tsinghua.edu.cn

Out-of-plane spin polarization $\sigma_z$ has attracted increasing interests of researchers recently, due to its potential in high-density and low-power spintronic devices. Noncollinear antiferromagnet (AFM), which has unique 120° triangular spin configuration, has been discovered to possess $\sigma_z$. However, the physical origin of $\sigma_z$ in noncollinear AFM is still not clear, and the external magnetic field-free switching of perpendicular magnetic layer using the corresponding $\sigma_z$ has not been reported yet. Here, we use the cluster magnetic octupole in antiperovskite AFM $Mn_3SnN$ to demonstrate the generation of $\sigma_z$. $\sigma_z$ is induced by the precession of carrier spins when currents flow through the cluster magnetic octupole, which also relies on the direction of the cluster magnetic octupole in conjunction with the applied current. With the aid of $\sigma_z$, current induced spin-orbit torque (SOT) switching of adjacent perpendicular ferromagnet is realized without external magnetic field. Our findings present a new perspective to the generation of out-of-plane spin polarizations via noncollinear AFM spin structure, and provide a potential path to realize ultrafast high-density applications.

**Introduction**

To realize non-volatile magnetic memories with the advantages of high density, high speed, and low power consumption, current induced spin-orbit torque (SOT) has been investigated due to its potentials in efficient and ultrafast manipulation of the magnetization in magnetic random access memory (MRAM) and logic devices by electrical methods[1-5]. For a perpendicularly magnetized ferromagnet (FM), which is preferred for high-density memories, the current-induced SOT switching usually needs an external magnetic field and a high critical current density (especially when the device is miniaturized), if the spin-source materials, such as heavy metals, could only provide a transverse in-plane spin polarization $\boldsymbol{\sigma}_y$ (along the y direction) by a longitudinal charge current (along the *x*-direction)[1,2,6,7]. To realize the external magnetic field-free switching of perpendicular magnetic layer, one effective approach is to use out-of-plane spin polarizations $\boldsymbol{\sigma}_z$ (along the z direction), which could also improve the switching efficiency[1,2,8,9]. Recently, there have been attempts aiming to obtain $\boldsymbol{\sigma}_z$, which derives from the spin-orbit scattering, filtering and spin precession of ferromagnetic interfaces or low-symmetric interfaces[9-13]. Considering practical device applications, using the long-range magnetic order within the bulk of the spin-source layer instead of optimizing the interface of different layers could be more uncomplicated and reliable[14].

Thanks to myriad magnetic and structural transitions, antiperovskite manganese nitrides $Mn_3AN$ (where A =Ga, Sn, Ni, etc., with a perovskite structure yet the cation and anion positions are interchanged), have attracted a revival of interest due to ample physical phenomena, such as negative thermal expansion, piezomagnetic, baromagnetic, and barocaloric effects[14-19]. Below their Néel temperature, the noncollinear antiferromagnetic (AFM) $Mn_3AN$ has been discovered to exhibit large anomalous Hall effect (AHE) due to the nonzero Berry curvature caused by the 120 ° triangular spin texture in the Kagome (111) plane[20-22], and the spin texture can be viewed as a ferroic ordering of a cluster magnetic octupole[23-25]. The 120 ° triangular spin texture could generate spin currents with out-of-plane spin polarization as well as corresponding spin torques. In Mn$_3$GaN films, **σ**$_z$ has been experimentally discovered[14], however, the relatively low Néel temperature ($T_N$ = 345 K) limits its practical device applications. More importantly, the physical origin of **σ**$_z$ in noncollinear AFM is still not clear, and the external magnetic field-free switching of perpendicular magnetic layer using **σ**$_z$ in noncollinear AFM has not been reported yet.

Here we consider another antiperovskite AFM Mn$_3$SnN with crystallographic structure shown in Fig. 1a, which has the highest Néel temperature ($T_N$ = 475 K) in the antiperovskite manganese nitride family[22,26]. **σ**$_z$ is generated by the precession of carrier spins when currents flow though the cluster magnetic octupole, and it is also dependent on the direction of the cluster magnetic octupole moment in conjunction with the applied current. **σ**$_z$ appears when the current **J** is parallel to the cluster magnetic octupole moment **T**, however, for the perpendicular case, **σ**$_z$ vanishes. The results are consistent with the magnetic symmetry analysis of noncollinear AFM spin texture where the magnetic mirror symmetry is broken[27]. Then, with the aid of **σ**$_z$, deterministic switching of a perpendicular magnet is realized even without an applied magnetic field.

**Results and discussion**

**Cluster magnetic octupole induced out-of-plane spin polarization.** Antiperovskite noncollinear AFM Mn$_3$SnN has the cubic structure $Pm\bar{3}m$ with finite magnetic moments of Mn in the Kagome-like (111) plane below the Néel temperature[20,22,26]. When we focus on the Mn atoms, the AFM $\Gamma_{4g}$ spin texture (Fig. 1b left) can be viewed as a ferroic ordering of a cluster magnetic octupole[20,22,26], which is composed of six Mn atoms (Fig. 1b right). The generation of AHE, anomalous Nernst effect and magneto-optical Kerr effect (MOKE) can be viewed as the emergence of the finite magnetization of the cluster magnetic octupole moment, like the case of $Mn_3Sn$[23-25,28]. As demonstrated in Fig. 1c, d, whether $\boldsymbol{\sigma}_z$ exists is decided by the direction of the cluster magnetic octupole moment **T** together with the applied current **J**. Fig. 1c illustrates a sketch map of the generation of $\boldsymbol{\sigma}_z$ where the carrier spins are rotated to out of plane by the spin-orbit field $\mathbf{H_{so}}$ (with the direction perpendicular to the applied current). This is similar to the case of FM (Ga,Mn)As[29] and collinear AFM $Mn_2Au$[8]. When **T** is (anti)parallel to **J**, the carrier spins (which are along **T**) is perpendicular to $\mathbf{H_{so}}$, a robust $\boldsymbol{\sigma}_z$ thus appears. On the contrary, when **T** is perpendicular to **J** as in Fig. 1d, the current-induced $\mathbf{H_{so}}$ is parallel to the carrier spins, which cannot trigger the spin precession and leads to the absence of $\boldsymbol{\sigma}_z$. $\boldsymbol{\sigma}_z$ can be expressed as

$$\boldsymbol{\sigma}_z \propto \mathbf{H_{so}} \times \mathbf{T}. \qquad (1)$$

Apart from $\boldsymbol{\sigma}_z$, $\boldsymbol{\sigma}_y$ can also be produced by the spin Hall effect (SHE) in $Mn_3SnN$, like the cases of IrMn and $Mn_3GaN$[14,27,30,31], which can produce SOT on the adjacent FM layer together with $\boldsymbol{\sigma}_z$.

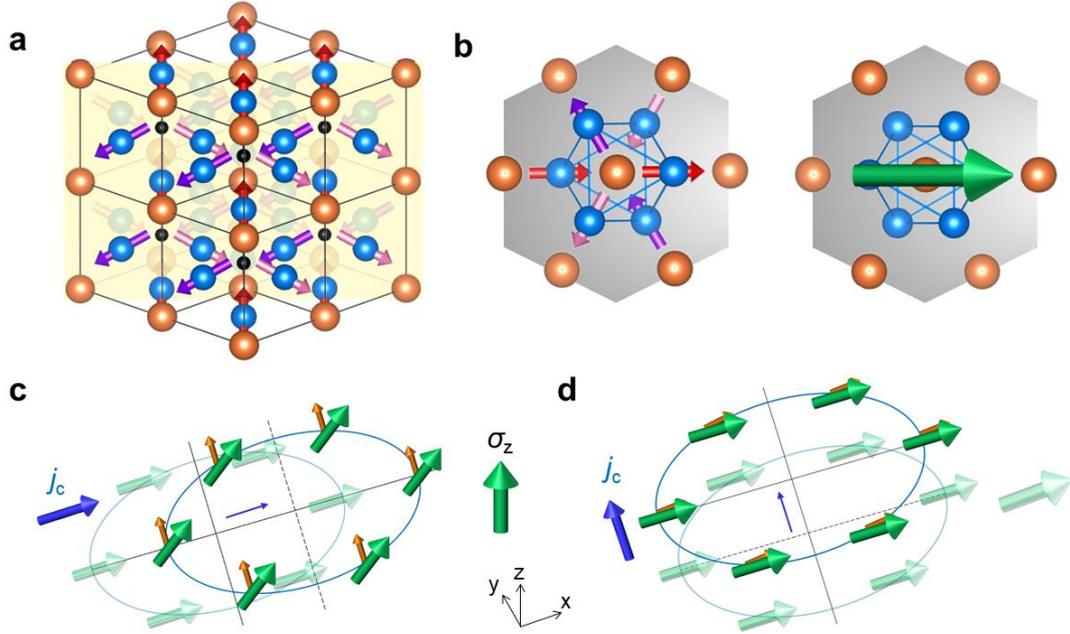

**Fig. 1** Generation of $\sigma_z$ in Mn$_3$SnN. **a** Crystal structure of Mn$_3$SnN, where the blue, orange and black spheres represent the Mn, Sn and N atoms, respectively. Red, purple and pink arrows denote magnetic moments of Mn atoms, and the yellow plane denotes the (110) plane. **b** Left, the spin structure of Mn$_3$SnN on the kagome bilayers, where the gray plane denotes the kagome (111) plane. Right, the ferroic ordering of a cluster magnetic octupole consisting of six spins viewed from the spin structure. The green arrow denotes the cluster magnetic octupole moment **T**. **c** Carrier spins are rotated by the spin-orbit field **H**$_{so}$ (brown arrows) when **J**//**T**, which induces the spin current with $\sigma_z$. **d** Spin rotation and $\sigma_z$ vanish when **J**⊥**T**. Here, **H**$_{so}$ is parallel to the carrier spins.

**Antiperovskite noncollinear AFM Mn$_3$SnN.** The Mn$_3$SnN films we used here is (110)-oriented, as revealed in Fig. 2a. The out-of-plane x-ray diffraction shows obvious peaks of Mn$_3$SnN (110) and (220), apart from the peak of the MgO substrate, indicating the quasi-epitaxial growth mode for the present Mn$_3$SnN films. The out-of-plane lattice constant is calculated to be 3.98 Å from the XRD pattern. Besides, there is no secondary phase within the sensitivity of XRD measurements, and the epitaxial growth is confirmed from the results of the Φ-scan measurement, as shown in Fig. 2b. The inspection of the Φ-scan shows that the peaks are separated by 180° with twofold symmetry for both the film and the substrate, indicating the crystallographic orientation relationship as MgO(110)[100]//Mn$_3$SnN(110)[100]. Through the energy dispersive spectrometer (EDS) and the X-ray photoelectron spectroscopy (XPS) quantitative analysis, the Mn:Sn:N atomic ratio of our film is 3:0.94:1.03, which is close to the nominal composition of Mn$_3$SnN. The surface morphology of the filmshows that the whole film is continuous and smooth, with the average surface roughness *Ra* being 0.199 nm. Fig. 2c presents the magnetic property of the 34 nm Mn$_3$SnN film with an out-of-plane magnetic field at 300 K. At a glance, the curve exhibits a diamagnetic behavior because of the diamagnetic background of the substrate, indicating the antiferromagnetic characteristic of the Mn$_3$SnN films. After the subtraction of the diamagnetic background, the film shows a small magnetization, which is also discovered in other noncollinear AFM films[21,32,33].

In the bulk Mn$_3$SnN, there may be two noncollinear AFM spin magnetic structures, $\Gamma_{5g}$ and $\Gamma_{4g}$. From the symmetric analysis of the Berry curvature, only the $\Gamma_{4g}$ rather than the $\Gamma_{5g}$ spin configuration could present AHE[21,32,33]. Fig. 2d presents the magnetic field dependent Hall resistivity $\rho_H$ of the single oriented Mn$_3$SnN film. The $\rho_H$ curve exhibits obvious AHE at both 300 K and 100 K, indicating that the magnetic structure of the film in our experiment is $\Gamma_{4g}$, which is consistent with the theoretical research[20].

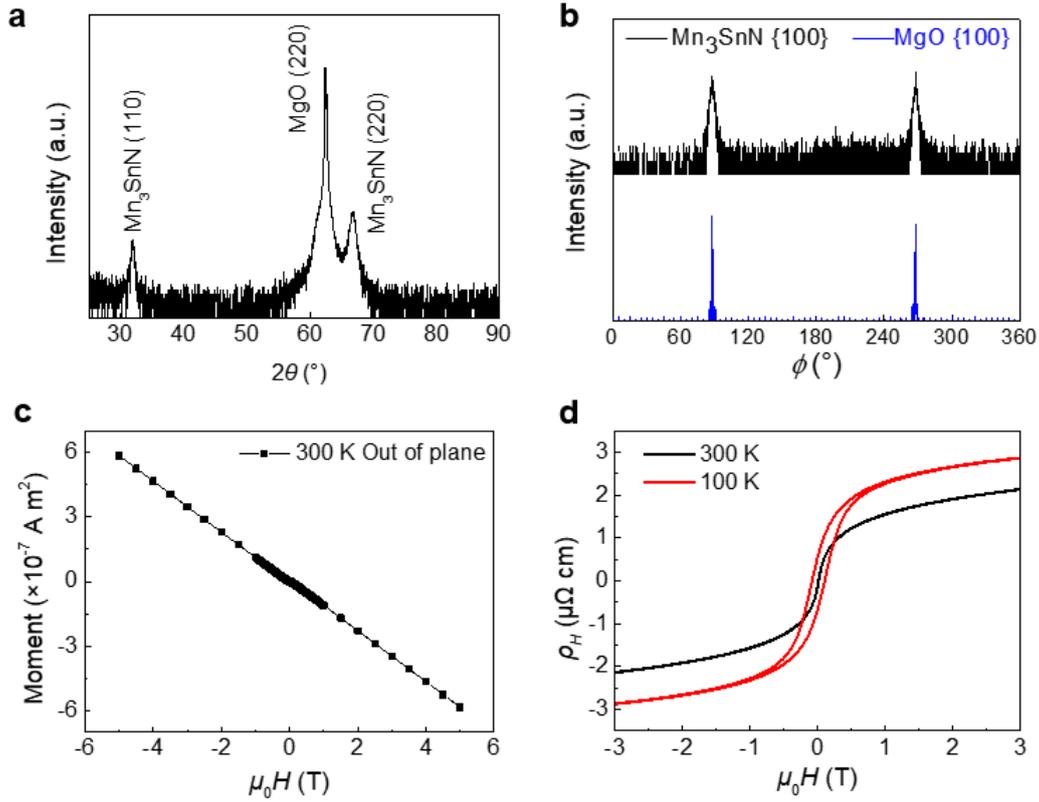

**Fig. 2** Basic properties of the Mn$_3$SnN film. **a** XRD patterns of the (110)-oriented Mn$_3$SnN films deposited on MgO (110) substrate. **b** $\Phi$ scan patterns of the {100} planes from the Mn$_3$SnN films and the MgO (110) substrate. **c** Magnetization hysteresis loops of the 34 nm Mn$_3$SnN (110) film deposited on MgO (110) substrate by an out-of-plane magnetic field at 300 K. **d** Magnetic field dependence of $\rho_H$ measured with an out-of-plane magnetic field at 300 K and 100 K, which exhibits obvious AHE in the film.

**Spin-torque ferromagnetic resonance measurement.** To evaluate the properties of the torques on the FM layer caused by the spin polarizations, we use Mn$_3$SnN(16 nm)/Py(12 nm) samples via the spin-torque ferromagnetic resonance (ST-FMR) technique[34]. For the ST-FMR measurement, as shown in Fig. 3a, a microwave (with high frequency of 2–17 GHz) current flowing through Mn$_3$SnN induces alternating torques on the Py later, and excites the magnetic moment of Py into precession. The process is detected as a resonant DC electrical signal $V_{mix}$ originated from the anisotropic magnetoresistance (AMR) of Py when the frequency of the microwave current and the in-plane applied magnetic field (with a field angle $\varphi$) meet the FMR condition[34-36]. When $\varphi$ is fixed at 45°, the data in Fig. 3b shows the relation of the frequency $f$ and the resonance position $H_0$, which can be well fitted by the Kittel formula, indicating that the resonance peak originates from the FMR of the Py layer and that the attached $Mn_3SnN$ film has no clear influence on the magnetic properties of Py layer[37].

The features of the torques can be recognized from the angular dependent line-shape of the resonance peaks (see Methods). Both in-plane and out-of-plane torque components can be obtained individually, since the symmetric ($V_S$) and antisymmetric ($V_A$) signals are proportional to the amplitude of the in-plane $\tau_\parallel$ and out-of-plane $\tau_\perp$ torque components, respectively[14,38,39]. Fig. 3c shows representative ST-FMR signals $V_{mix}$ of the device bar measured with the frequency of 5 GHz, power of 20 dBm and $\varphi$ of 100°. The microwave current is applied along [001] direction of the (110)-oriented $Mn_3SnN$ film, as shown in Fig. 3d. According to the established analysis of ST-FMR data, comparison of the ST-FMR signals measured at negative and positive magnetic fields could qualitatively reflect the influences of SOT triggered by $\boldsymbol{\sigma_z}$. Unlike the condition of $\boldsymbol{\sigma_y}$, the symmetric signals $V_S$ have the same signs and the antisymmetric signals $V_A$ have evidently different amplitude, presenting obvious evidence for a prominent contribution from SOT triggered by $\boldsymbol{\sigma_z}$[1,8,9].

To quantitatively recognize the torque components, we conduct the ST-FMR measurement with different in-plane magnetic field angle $\varphi$. According to the spin rectification theory of AMR[39], $V_S$ ($V_A$) depends on the product of the angular related AMR and $\tau_\parallel$ ($\tau_\perp$). Merely taking into account the conventional SHE or the Rashba-Edelstein effect (REE) and Oersted field, the conventional antidamping torque $\tau_S \propto \boldsymbol{m} \times (\boldsymbol{m} \times \boldsymbol{\sigma}_y)$ results in a symmetric line-shape with an amplitude of $V_S = S \sin 2\varphi \cos \varphi$, while the field-like torque $\tau_A$ induced by the Oersted field results in an antisymmetric line-shape with an amplitude of $V_A = A \sin 2\varphi \cos \varphi$, where S and A are constant terms. Hence, $V_S$ and $V_A$ exhibit the same angular dependence for a conventional HM/FM heterostructure[9,14,34]. In Fig. 3e, f, however, we find that the angular dependence of $V_S$ and $V_A$ for Mn$_3$SnN is obviously different from this simple case, which can be fitted by adding additional torque terms with the presence of $\boldsymbol{\sigma}_z$: $\tau_B \propto \boldsymbol{m} \times \boldsymbol{\sigma}_z$, (the in-plane field-like torque induced by $\boldsymbol{\sigma}_z$), and $\tau_C \propto \boldsymbol{m} \times (\boldsymbol{m} \times \boldsymbol{\sigma}_z)$, (the out-of-plane antidamping torque induced by $\boldsymbol{\sigma}_z$)[9,14,34].

As explained above, $\boldsymbol{\sigma}_z$ in Mn$_3$SnN is generated by the noncollinear 120° triangular spin structure, and when the cluster magnetic octupole moment **T** is perpendicular to the applied current **J**, $\boldsymbol{\sigma}_z$ cannot be induced. Therefore $\boldsymbol{\sigma}_z$ is dependent on the measurement configuration. Specific 2D materials like WTe$_2$ have been discovered to have $\boldsymbol{\sigma}_z$ as a result of the broken crystal mirror symmetry ($M$)[9,40,41]. Here, in noncollinear AFM Mn$_3$SnN, although the crystal symmetry is maintained, $\boldsymbol{\sigma}_z$ exists due to the broken magnetic mirror symmetry ($M'$) which contains a time reversal symmetry $T$ ($M' = M*T$), like the case of IrMn[27,42]. We then consider the magnetic symmetry of Mn$_3$SnN. Fig. 3g shows another measurement configuration of ST-FMR at the same sample, where the microwave current is applied along [1$\bar{1}$0] direction, which is perpendicular to **T** as well as the magnetic mirror plane. From Fig. 3h, i, we can see that there is no clear $\boldsymbol{\sigma}_z$ in this way, indicating that $\boldsymbol{\sigma}_z$ is restricted by the symmetry relationship and related with the spin structure.

Indeed, for (110)-orientated Mn$_3$SnN with $\Gamma_{4g}$ magnetic configuration, the (110) plane is the magnetic mirror plane, and the cluster magnetic octupole moment **T** is also located in the plane. According to the analysis of magnetic asymmetry, when the current is applied along [001]-axis of Mn$_3$SnN, it is parallel to the magnetic mirror plane, with an angle between the current and **T** being 35 degree, so **σ$_z$** induced $\tau_B$ and $\tau_C$ are allowed[27]. Consequently, **σ$_z$** shows up. The generation of the spin torque relative to the charge current density can be parameterized into the spin-torque ratio[14] (See Methods). The antidamping and field-like spin torque ratios of **σ$_z$**, $\theta_{AD,z}$ and $\theta_{FL,z}$ are 0.003±0.001, and 0.053±0.005, respectively. $\theta_{FL,z}$ of Mn$_3$SnN is larger than that of WTe$_2$ and MnPd$_3$, reflecting that **σ$_z$** mainly contributes to the field-like torque in Mn$_3$SnN/Py system. The comparatively large field-like torque is most likely due to the spin accumulation at the Mn$_3$SnN/Py interface, which interacts with the adjacent Py layer, producing an exchange field[8]. On the contrary, when the current is applied perpendicular to the magnetic mirror plane (also **T**), **σ$_z$** disappears accordingly. We find that the mechanism is also applicable to the (001)-oriented Mn$_3$SnN film by performing identical ST-FMR measurement of (001)-oriented Mn$_3$SnN(16 nm)/Py(12 nm) sample at room temperature. Moreover, ST-FMR measurements show that **σ$_z$** still exists at 380 K, but the antidamping and field-like spin torque ratios of **σ$_z$**, $\theta_{AD,z}$ and $\theta_{FL,z}$, are smaller than their counterparts at room temperature. These results verify that **σ$_z$** is closely related to the cluster magnetic octupole and the magnetic asymmetry of Mn$_3$SnN.

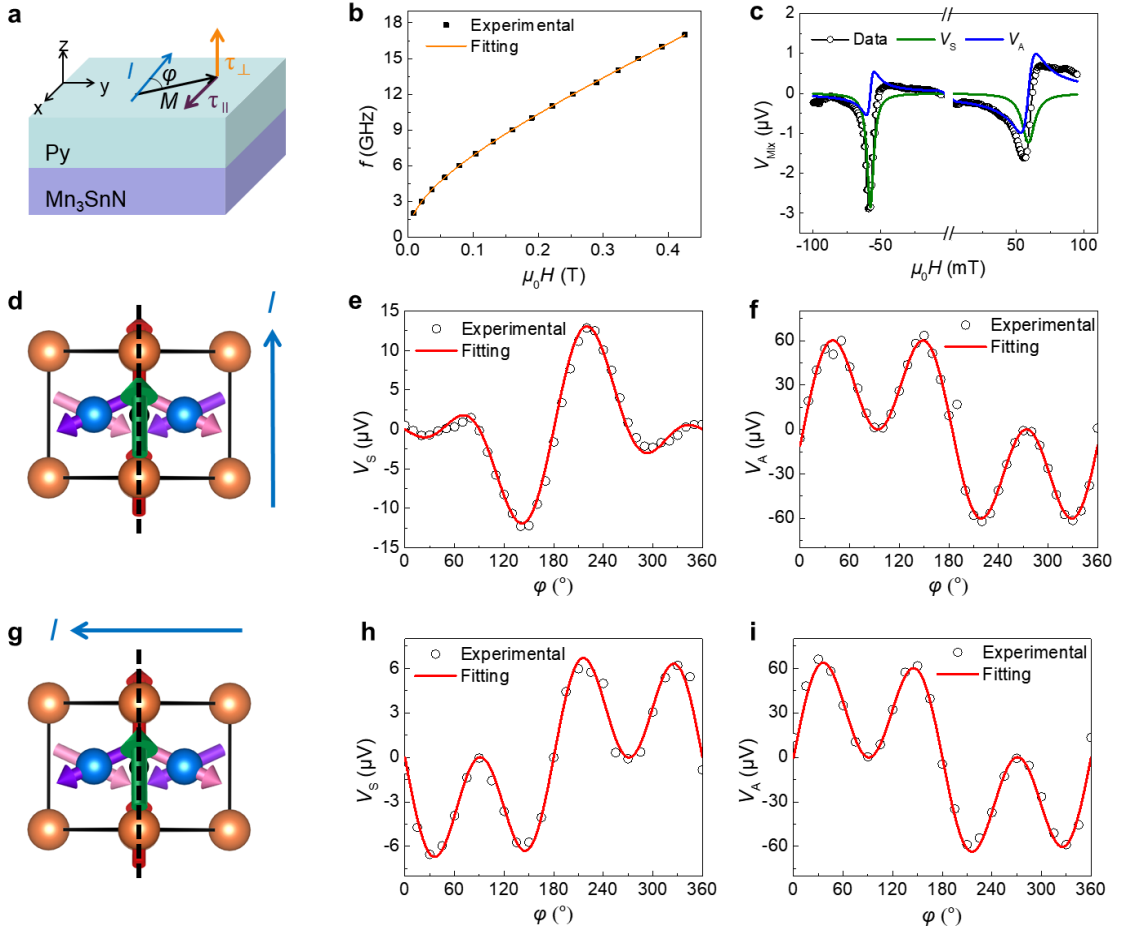

**Fig. 3** Current direction dependent $\sigma_z$ in (110)-orientated $Mn_3SnN$ device. **a** Schematic diagram of the ST-FMR geometry for $Mn_3SnN(110)$/Py bilayers. $\tau_\parallel$ and $\tau_\perp$ denote the in-plane and out-of-plane torque components. **b** The relationship of the frequency $f$ and the resonance position $H_0$, which can be well fitted by the Kittel formula. **c** ST-FMR signals $V_{mix}$ of the device bar measured with the frequency of 5 GHz, power of 20 dBm and $\varphi$ of 100° when the current channels are along the [001] direction. **d** Schematic diagram of magnetic structure and current direction. The current is applied parallel to the magnetic mirror plane (represented by the dashed black line). Angular dependence of line shape amplitude of ST–FMR signals for **e** symmetric and **f** antisymmetric signal in $Mn_3SnN$/Py structure. **g** The schematic diagram for the case when the current direction is perpendicular to the magnetic mirror plane. Angular dependence of line shape amplitude of ST–FMR signals for **h**

symmetric and **i** antisymmetric signal in the same sample.

**Magnetic field-free SOT switching.** Next, using the observed $\sigma_z$, let us discuss the SOT switching of a perpendicularly magnetized FM layer deposited on top of the Mn$_3$SnN film, with the structure of the sample being MgO/Mn$_3$SnN(12 nm)/(Co(0.4 nm)/Pd(0.8 nm))$_3$ stack. Fig. 4a shows the measurement configuration of the anomalous Hall resistance $R_{AHE}$ by applying a pulse current $I$ along the [001] direction (x-direction), where $\sigma_z$ exits. The pulse width is 1 ms, followed by a read current of 0.1 mA. The hysteresis R$_{AHE}$ loop with out-of-plane magnetic field in Fig. 4b confirms that perpendicular magnetic anisotropy (PMA) is present in the Co/Pd multilayer. Fig. 4c shows the current-induced field-free SOT magnetization switching of Co/Pd multilayer caused by $\sigma_z$, with a hysteretic behavior and a sign change in the absence of applied magnetic field. The current density required to achieve the field-free SOT switching of our Mn$_3$SnN/(Co/Pd)$_3$ sample is estimated to be approximately $9 \times 10^6$ A cm$^{-2}$. The present current density is comparable to that of traditional field-free switching by wedged structure or exchange bias[43,44]. High crystal quality Mn$_3$SnN films with a highly ordered magnetic configuration or other noncollinear AFM are highly warranted for stronger $\sigma_z$, to decrease the current density for the $\sigma_z$-induced field-free switching. The corresponding MOKE microscope images of the field-free SOT magnetization switching under the condition (i)-(iii) are shown in Fig. 4g-i. Apparently, the magnetization of Co/Pd multilayer can be switched by the positive and negative currents without external magnetic field, and switching only occurs in the current channel, with the magnetization in the voltage channel maintaining the original state. The MOKE images directly reveal that the resistance change in Fig. 4c is caused by the current induced SOT magnetization switching rather than other effects[8,45,46]. The SOT switching does not decline after cycling 10 times (Supplementary Fig. S6), reflecting the robustness of the device. From the $R_{AHE}$-$I$ loop, we can see that approximate 60% of the magnetic Co/Pd multilayer volume switches, compared with the saturation AHE resistance obtained in the $R_{AHE}$-$H$ loop. Combined with the MOKE figure, we owe this phenomenon to the following reason. For the magnetic field-induced switching, the whole area of the Co/Pd multilayer in the cross pattern switches, including the Hall leads. But for the current-induced switching, only the current path switches while the Hall leads do not switch, as shown in Fig. 4h. Considering how the Hall leads detect the AHE signal, the current-induced switching yields a smaller AHE voltage than the magnetic field-induced case.

The in-plane $M$-$H$ curve of Mn$_3$SnN/(Co/Pd)$_3$ shows no exchange bias at room temperature, revealing that the field-free switching is irrelevant to the exchange bias[8]. Fig. 4d and Fig. 4e exhibit the measured $R_{AHE}$-$I$ loops for negative and positive magnetic fields of 50 mT applied along the [001] direction, respectively. Obviously, the switching polarity is reversed upon reversing the field to opposite direction, which is consistent with the typical SOT switching of a perpendicularly magnetized FM[1,2,5]. When we apply a large magnetic field of 500 mT along the [001] direction, which is larger than the anisotropy field of the Co/Pd multilayer, there is no switching signals, showing that the variation of anomalous Hall resistance is not triggered by the thermal effects caused by the pulse currents[1,2,17,47]. To confirm whether the $\sigma_z$ generated by the noncollinear antiferromagnetic configuration in Mn$_3$SnN is responsible for the observed field-free switching, we apply the current along the [1$\bar{1}$0] direction ($y$-direction), where no $\sigma_z$ exists according to the results of ST-FMR measurement. As shown in Fig. 4f, the measured $R_{AHE}$-$I$ curve does not show obvious hysteretic behavior, revealing that $\sigma_z$ is the main reason for the field-free switching of the magnetic moments of Co/Pd multilayer. The direction related field-free SOT switching here also illustrates that the weak magnetization of the film does not have obvious influence on the switching measurement.

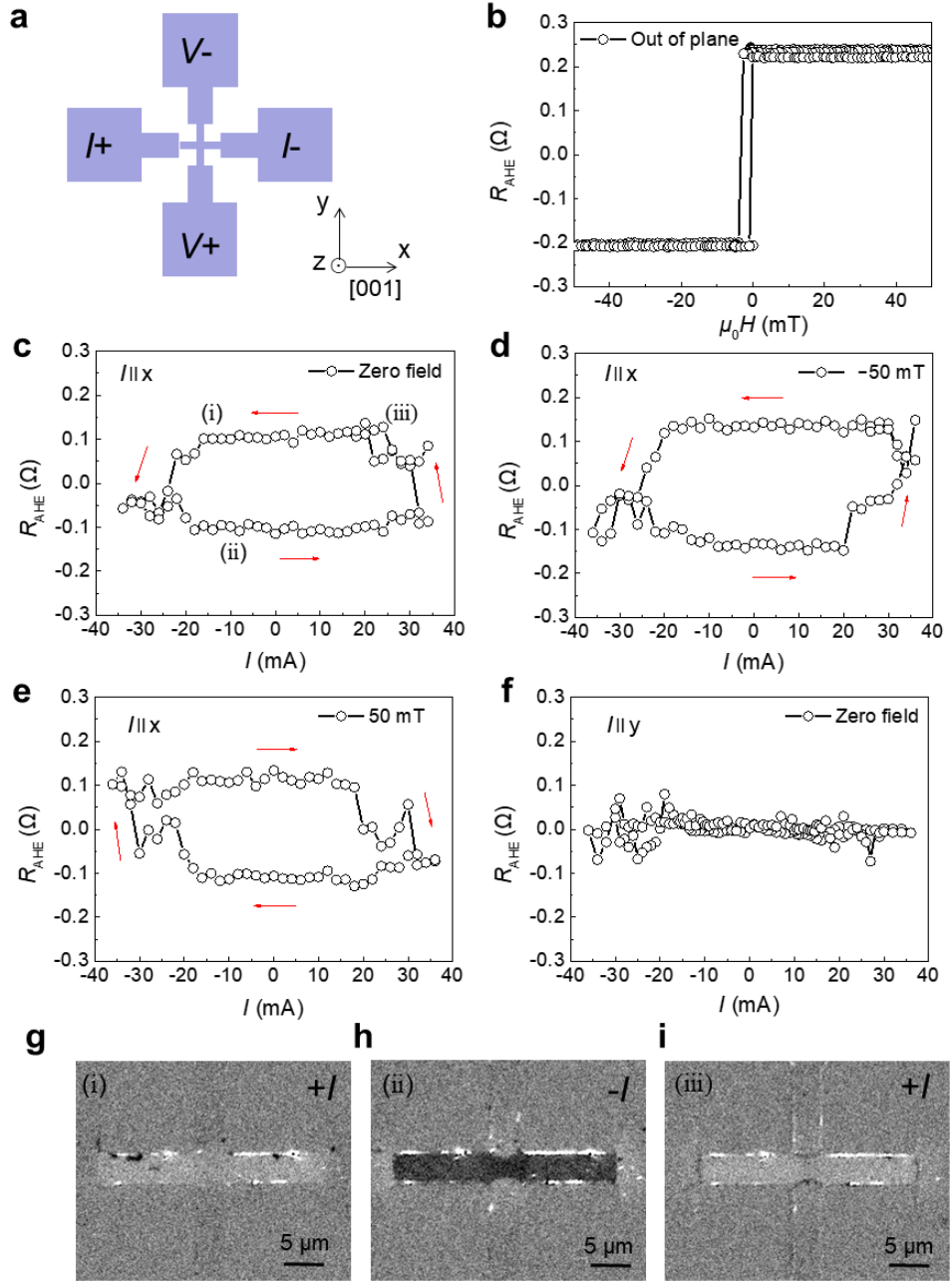

**Fig. 4** Current induced SOT switching in $Mn_3SnN/(Co/Pd)_3$ stacks. **a** The schematic diagram of the device using for electrical transport measurements. **b** The anomalous Hall resistance $R_{AHE}$ as a function of the out-of-plane external field. **c** Current-induced

field-free SOT switching shown by $R_{AHE}$ as a function of current magnitudes at zero field with current along *x*-direction. Current-induced SOT switching under the application of –50 mT **d** and 50 mT **e** magnetic field along x-direction, respectively. **f** $R_{AHE}$ as a function of current magnitudes at zero field with current along y-direction. **g-i** MOKE images of the stack films, which imprint the out-of-plane magnetization of the Co/Pd multilayer corresponding to the condition of (i) to (iii) in **c**.

**Conclusion**

In conclusion, we have demonstrated the origin of out-of-plane spin polarization $\boldsymbol{\sigma}_z$ in noncollinear AFM $Mn_3SnN$ at room temperature. $\boldsymbol{\sigma}_z$ is induced by the precession of carrier spins when currents flow though the cluster magnetic octupole, which also relies on the direction of the cluster magnetic octupole in conjunction with the applied current. The field-free SOT switching of a perpendicularly magnetized FM is then realized in $Mn_3SnN/(Co/Pd)_3$ stacks with the aid of $\boldsymbol{\sigma}_z$. In addition to $Mn_3SnN$, the out-of-plane spin polarization is expected to exist in other noncollinear AFM systems with 120° triangular spin texture, and can be used to switch the magnetic moments of FM efficiently without external magnetic field. Our findings enrich the current comprehension of spin-current physics and provides potential chances to use $\boldsymbol{\sigma}_z$ in the next generation of SOT-based spintronics devices.

**Methods**

**Sample preparation.** The (110)-oriented $Mn_3SnN$ films we used here were deposited on MgO (110) substrates by magnetron sputtering. The optimized growth temperature was 400 °C. The base pressure was $3\times10^5$ Pa and the process gas (with 13% $N_2$ of Ar) pressure was 0.5 Pa with the growth rate being 0.08 nm/s. Considering the different sputtering yields of Mn and Sn elements as well as the fact that the compound of Mn and Sn was stable only in the presence of excess Mn, a $Mn_{81}Sn_{19}$ target was used for the film growth[48]. Py or perpendicularly magnetized Co/Pd multilayer was then deposited on $Mn_3SnN$ by magnetron sputtering at room temperature.

**Sample characterization.** X-ray diffraction (XRD) and XRR of the Mn3SnN films was measured using Cu Kα1 radiation with λ = 1.5406 Å. The surface roughness was characterized by atomic force microscope (AFM). Magnetic properties were measured by a superconducting quantum interference device (SQUID) magnetometry with the field up to 5 Tesla. The magnetotransport measurements were conducted using a physical property measurement system (PPMS). The magnetization reversal images were obtained using magneto-optical Kerr microscopy (MOKE).

**Device fabrication.** To measure the SOTs in $Mn_3SnN$/Py sample using ST-FMR technology, the thin films were patterned into microstrips devices along different crystallographic directions with the size of 30 μm × 20 μm by standard photolithography and Ar-ion milling techniques. Top electrodes of Ti (10 nm)/Pd (50 nm) were then deposited by e-beam evaporation. ST-FMR measurements were conducted by injecting currents into the device with frequency from 2 to 17 GHz. A magnetic field $H$ was applied in the sample plane and at an angle $\varphi$ to the current direction. To conduct the SOT switching experiment, $Mn_3SnN/(Co/Pd)_3$ samples were patterned into cross bar devices with the channel width being 5 μm. The current pulses had a constant duration of 3 s but varying amplitude. The current pulse width is 1 ms, followed by a read current of 0.1 mA.

**ST-FMR analysis.** The relation of the frequency $f$ and the resonance position $H_0$

is fitted by the Kittel formula, which is expressed as $f = \frac{\gamma}{2\pi}[H_0(H_0 + M_{\text{eff}})]^{1/2}$, where $M_{\text{eff}}$ and $\gamma$ represent the effective magnetization and the gyromagnetic ratio, respectively. Using the line-shape fitting equation:

$$V_{mix}(H) = V_S \frac{\Delta H^2}{\Delta H^2 + (H-H_0)^2} + V_A \frac{\Delta H(H-H_0)}{\Delta H^2 + (H-H_0)^2}, \qquad (2)$$

both in-plane (the first term) and out-of-plane (the second term) torque components can be obtained individually. The symmetric ($V_S$) and antisymmetric ($V_A$) amplitude of the Lorentzian line shape are proportional to the amplitude of the in-plane $\tau_\parallel$ and out-of-plane $\tau_\perp$ torque components, respectively, with the following relationship:

$$V_S = -\frac{I_{\text{rf}}}{2}\left(\frac{dR}{d\varphi}\right)\frac{1}{\alpha(2\mu_0 H_0 + \mu_0 M_{\text{eff}})}\tau_\parallel \qquad (3)$$

$$V_A = -\frac{I_{\text{rf}}}{2}\left(\frac{dR}{d\varphi}\right)\frac{\sqrt{1+M_{\text{eff}}/H_0}}{\alpha(2\mu_0 H_0 + \mu_0 M_{\text{eff}})}\tau_\perp \qquad (4)$$

where, $\Delta H$ is the width of the resonance peak, $I_{\text{rf}}$ is the microwave current, $R$ is the resistance as a function of the in-plane magnetic field angle $\varphi$ due to the AMR of Py, and $\alpha$ is the Gilbert damping coefficient. According to the spin rectification theory of AMR, $V_S$ ($V_A$) depends on the product of the angular related AMR and $\tau_\parallel$ ($\tau_\perp$). Taking into account the presence of $\sigma_z$, $V_S$ and $V_A$ can be calculated as:

$$V_S = S \sin 2\varphi \cos \varphi + B \sin 2\varphi \qquad (5)$$

$$V_A = A \sin 2\varphi \cos \varphi + C \sin 2\varphi \qquad (6)$$

where $S$, $B$, $A$ and $C$ are constant terms. When we leave out the AMR rectification ($\sin 2\varphi$), the angular dependencies of the in-plane and perpendicular torque amplitudes are:

$$\tau_\parallel(\varphi) = \tau_S \cos \varphi + \tau_B \qquad (7)$$

$$\tau_\perp(\varphi) = \tau_A \cos \varphi + \tau_C \qquad (8)$$

where $\tau_S$, $\tau_B$, $\tau_A$, and $\tau_C$ are values independent of $\varphi$, where $\tau_S$ represents the contribution from the antidamping torque induced by $\boldsymbol{\sigma_y}$; $\tau_A$ represents the contribution from the torque of the current-induced Oersted field; $\tau_B$ represents the field-like torque induced by $\boldsymbol{\sigma_z}$; $\tau_C$ represents the antidamping torque induced by $\boldsymbol{\sigma_z}$, respectively. The spin torque ratios are calculated as follows:

$$\theta_{\mathrm{AD},z} = \frac{\tau_C}{\tau_A} \frac{e\mu_0 M_S t_{\mathrm{Py}} t_{\mathrm{Mn_3SnN}}}{\hbar} \tag{9}$$

$$\theta_{\mathrm{FL},z} = \frac{\tau_B}{\tau_A} \frac{e\mu_0 M_S t_{\mathrm{Py}} t_{\mathrm{Mn_3SnN}}}{\hbar} \tag{10}$$

where $e$ is the electron charge, $M_S$ is the saturation magnetization of Py (which can be replaced by the effective magnetization determined by ST-FMR), $t_{\mathrm{Py}}$ is the thickness of Py, $t_{\mathrm{Mn_3SnN}}$ is the thickness of Mn$_3$SnN, and $\hbar$ is the reduced Planck's constant[9,14,27,34,36,39].

**Note added in proof**

After finishing the current work, we became aware of two relevant works that demonstrated the SOT switching using Mn$_3$Sn[49] and MnPd$_3$[50] as the spin source. In Mn$_3$Sn, field-free SOT switching of perpendicular magnetic layer is realized by the magnetic spin Hall effect, while in MnPd$_3$, it is caused by the out-of-plane spin polarization $\boldsymbol{\sigma_z}$ induced by the low symmetry of the (114)-oriented MnPd$_3$ thin films. Our work uses the cluster magnetic octupole in antiperovskite AFM Mn$_3$SnN to demonstrate the generation of $\boldsymbol{\sigma_z}$, and realizes the field-free SOT switching of perpendicular magnetic layer by $\boldsymbol{\sigma_z}$ generated in Mn$_3$SnN. Moreover, we find that whether $\boldsymbol{\sigma_z}$ exists in noncollinear AFM relies on the direction of the cluster magnetic octupole **T** in conjunction with the applied current **J**.

**Acknowledgements**

The National Natural Science Foundation of China (Grant No. 51871130), and the Natural Science Foundation of Beijing, China (Grant No. JQ20010). We are grateful to the support of ICFC, Tsinghua University.